\documentclass[articlesize,journal]{IEEEtran}
\usepackage{amsmath,amsfonts}
\usepackage{algorithmic}
\usepackage{algorithm}
\usepackage{tabularx}
\usepackage{array}
\usepackage[caption=false,font=normalsize,labelfont=sf,textfont=sf]{subfig}
\usepackage{textcomp}
\usepackage{stfloats}
\usepackage{url}
\usepackage{siunitx}
\usepackage{commath,amsmath}
\usepackage{mathtools}
\usepackage{derivative}\usepackage{booktabs,tabularx}
\usepackage{verbatim}
\usepackage{graphicx}
\usepackage{cite}
\begin{document}

\title{Atomic Anatomy of Low-Inertia Power Systems}

\author{Subham Sahoo, \textit{Senior Member, IEEE}, Arpan Malkhandi and Kristian Skafte Jensen
\thanks{This work is supported by the Nordic Energy Research programme via Next-uGrid project n. 117766.}}



\maketitle

\begin{abstract}
In this letter, we determine a fundamental anatomical modeling parallelism between low-inertia power systems and Bohr's atomic model. The proposed atomic architecture will serve as a \textit{microscopic building block}, where we validate the structural analogy of low-inertia power systems using semi-classical quantum approximations in IEEE 9-bus system. As a future scope of work, detailed modeling \& system stability will be investigated using atomic physics based principles.
\end{abstract}

\begin{IEEEkeywords}
Low-inertia power systems, Bohr model, Semi-classical models, Power system stability, Power systems modeling.
\end{IEEEkeywords}
\section{Introduction}
\IEEEPARstart{I}{nertia} in power systems is usually governed by the inherent rotational mass in synchronous machines (SM) \cite{dirk}. From a system perspective, converter-interfaced generation (CIG) that allow integration of renewable energy sources behave quite differently from SM. Apart from their intermittent nature, CIG without any stored kinetic energy do not \textit{implicitly} contribute to the system inertia due to electrical decoupling. With increased penetration of power from renewable energy sources and the current decarbonization goals, the future power network is highly prone to system instability.

From a curiosity-driven research viewpoint, this letter proposes a fundamentally new anatomical modeling parallelism between the low-inertia power systems and Bohr's atomic model by hypothesizing a structural duality of the system with respect to different components in the smallest fundamental particle, i.e., an atom. A transformative understanding to model future power systems as an atom will not only unravel generation of better set-points for reduced numerical computational efforts to assess the system stability, but will also serve as a building block based planning tool to encompass power system expansion and integration in the future only using inter-atomic physical interactions. We argue against the single mass model of power systems that represent frequency as a global parameter by splitting it spatially into the rotational inertia in SM and \textit{virtual} inertia in CIG \cite{vsg} (commonly known as virtual synchronous generators (VSGs)) of the future power systems into different atomic components as per Bohr's prophecy. Finally, we validate the proposed structural duality based on the center-of-mass theory and semi-classical quantum approximations in Bohr's model.

\section{The Bohr Atomic Model}
\begin{figure}[!t]
\centering
\includegraphics[width=\columnwidth]{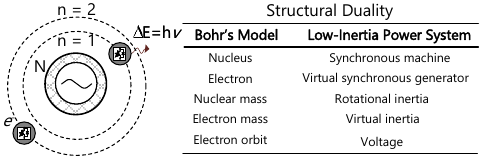}
\caption{The atomic solar system -- Atomic nucleus depicting a SM and two of the possible orbits for an electron, denoting a VSG. When the activation energy $\Delta E$ is released by an electron, it moves from one orbit to another -- similar to how the terminal voltage across a VSG changes under a step change in reference energy.}
\label{fig:structure}
\end{figure}

Proposed by Niels Bohr in his paper “\textit{On the Constitution of Atoms and Molecules}”\cite{Bohr_1913} in 1913, the Bohr atomic model is a structural representation of a hydrogen atom consisting of a nucleus made up of a single proton, and an orbiting electron, as shown in Fig. \ref{fig:structure}. The model is best summarized by three fundamental postulates:
\begin{enumerate}
    \item As opposed to the classical theory of electromagnetism, electron(s) can be located at certain stable orbits, known as stationary orbits, without either absorbing or emitting any energy. 
    \item These stationary orbits are attained at discrete distance from the nucleus. Specifically, orbits are attained at precise, unchangeable radii, wherein the angular momentum of the orbiting electron is an integer multiple of the reduced Planck constant, $\hbar$ in the quantization equation, $L=n\hbar$, where $n$ is the principal quantum number and describes the energy level of the electron.
    \item Electrons can only interact using electromagnetic radiation by jumping from one orbit to another. An electron can jump to a higher energy level by absorbing a photon, and can jump to a lower energy level by emitting a photon. The energy of these photons is determined as the difference between the energies of the two orbits, which by the Planck relation gives $\Delta E = E_f - E_i = h\nu$, where $E_f$ and $E_i$ are the energies of the final and initial orbits, respectively. Furthermore, $h$ is the Planck constant, and $\nu$ is the frequency of the photon.
\end{enumerate}

\textit{\textbf{Remark I:} The energy dissipation/absorption $\Delta E$ in Bohr's model is a spatial phenomena with its bounded orbital radius scattered around the nucleus.}
\section{Structural Duality}
\subsection{Mass Identification in Systems \& Converters}
With system frequency considered as a global parameter, the SMs can be aggregated into a single-mass model with its rotor dynamics given by: \begin{equation}\label{eq:swing}
    \od[2]{\delta}{t} = \frac{\omega_s}{2H}(P_m - P_e),
\end{equation}
where, $\delta$ is the electric angle of the SM. Further, $P_m$ and $P_e$ are mechanical and electrical power, respectively, with $\omega_s$ being the synchronous speed. Lastly, $H = \frac{J\omega_s^2}{2S_{\text{rated}}}$ is the \textit{normalized inertia constant}, where $J$ is the moment of inertia and $S_{\text{rated}}$ is the rated power of SM. 
\begin{table}[!thb]
\centering
\caption{Structural Insights from IEEE 9-Bus System}
\begin{tabular}{c}
\textbf{1. Case study with 1 VSG \& 2 SMs}
\end{tabular}
\resizebox{\columnwidth}{!}{%
\begin{tabular}{@{\extracolsep{\fill}}ccccccc}
\toprule
Case & \textbf{VSG} & \textit{V}$^a$, \textit{P}$^b$ & \textbf{SM1} & \textit{V}, \textit{P} & \textbf{SM2} & \textit{V}, \textit{P} \\
\midrule
Case I & \#2*                    & 1.006,140 & \#3 & 0.997,170 & \#1 & 0.993,170 \\
Case II & \#6                   & 0.984,133 & \#3 & 0.971,174 & \#1 & 0.976,174 \\
Case III & \#1 & 0.983,127 & \#2 & 0.963,172 & \#3 & 0.963,172 \\
\bottomrule
\end{tabular}}
\\
\begin{tabular}{c}
\textbf{2. Case study with 2 VSGs \& 1 SM}
\end{tabular}
\resizebox{\columnwidth}{!}{%
\begin{tabular}{@{\extracolsep{\fill}}ccccccc}\toprule
Case & \textbf{VSG1} & \textit{V}, \textit{P} & \textbf{VSG2} & \textit{V}, \textit{P} & \textbf{SM} & \textit{V}, \textit{P} \\
\midrule
Case IV & \#1                    & 1,156 & \#2 & 0.996,125 & \#3 & 1.03,212 \\
Case V & \#3                   & 1.004,152 & \#2 & 1.001,121 & \#1 & 0.979,214 \\
\bottomrule
\end{tabular}}
\\
\begin{tabular}{c}
\textbf{3. Case study with 3 SMs/VSGs}
\end{tabular}
\resizebox{\columnwidth}{!}{%
\begin{tabular}{@{\extracolsep{\fill}}ccccccc}\toprule
Case & \textbf{\#1} & \textit{V}, \textit{P} & \textbf{\#2} & \textit{V}, \textit{P} & \textbf{\#3} & \textit{V}, \textit{P} \\
\midrule
Case VI & SM1                    & 0.978,148 & SM2 & 0.96,132 & SM3 & 0.984,164 \\
Case VII & VSG1                   & 0.998,143 & VSG2 & 0.994,165 & VSG3 & 0.986,198 \\
Case VII & VSG1                  & 0.995,148 & VSG2 & 0.988,176 & VSG3 & 1.012,196 \\
\bottomrule
\end{tabular}}
\footnotesize{$^*$ denote the bus number, $^a$$V$ is normalized in p.u., $^b$$P$ is stated in MW.}
\label{tab:equivalence}
\end{table}



For a solid cylinder rotating about an axis (such as a rotor shaft in SM), the moment of inertia $J = \frac{1}{2}mr^2$, where $m$ and $r$ denote the mass and radius of the cylinder, respectively. Hence, a higher mass not only contributes to a higher moment of inertia, but also naturally align to a smaller change in frequency during contingencies. 
On the other hand, VSGs do not have any inherent physical mass, and thus do not benefit from natural storage of kinetic energy. However, the maximum amount of energy stored in its DC capacitor, given by $E_C = \frac{1}{2}CV^2$ with $C$ being the DC capacitance and $V$ being the DC voltage across it, is programmed to provide \textit{virtual inertial} response similar to the mechanical inertia. 

\textit{\textbf{Remark II:} Comparing the expressions for moment of inertia $J$ and stored energy in capacitors $E_C$ (with $C$ as the equivalent mass in VSGs), an important disparity is the scale of the characteristic mass between SMs (rotor) \& VSGs (DC capacitence) with $m\gg m_e$.} {It is worthy notifying that this letter consistently follows the definition of inertia of SMs and VSGs in \cite{dirk}.}

\textit{\textbf{Corollary I:} Using the mass disparity in Remark II, single mass $m$ of the system can be visualized as the nucleus in Bohr's model, which accounts for majority of the atomic mass, whereas the electrons (analogous to the VSGs) orbiting around it signify negligible system mass $m_e$.}
\subsection{Center of Mass for Particle Systems}
Considering each SM in power systems as a particle, we will then have distributed mass corresponding to the rotor of each SM.

\textit{\textbf{Hypothesis:} In low-inertia power systems with distributed mass, the center of mass can represent the position of the atomic nucleus, and consequently, can attribute the position of orbits around it.}

In that regard, we consider a very \textit{naive} approach of a system of particles $P_i$ with $i=1,2,\ldots,n$, where the particles have mass $m_i$, and are located in a space with coordinates ${r}_i$. Finally, the coordinates for the center of mass $\mathbf{R}$, satisfy:
\begin{equation}\label{eq:com}
    \sum_{i=1}^n m_i({r}_i-\mathbf{R}) = \mathbf{0}.
\end{equation}
We can obtain $\mathbf{R}$ simply by isolating:
\begin{equation}\label{eq:r}
    \mathbf{R} = \frac{1}{M}\sum_{i=1}^n m_i{r}_i
\end{equation}
where, $M=\sum_{i=1}^n m_i$ is the total mass of the system. 

\textit{\textbf{Remark III:} The typical definition of ${r}_i$ in Bohr's model does not necessarily extend to low-inertia power systems since the former caters to a spatial phenomena as per Remark I, where the activation energy $\Delta E$ required by an electron to jump from one orbit to another can be absorbed/dissipated anywhere in the space. On the contrary, energy dissipation in power systems is only possible across the transmission lines, which necessitates a viable definition of ${r}_i$ in low-inertia power systems.}

Conferring on key structural parallelism to this end, a close representation of the atomic anatomy of low-inertia power systems can be seen in Fig. \ref{fig:structure}. However, there are still some unresolved questions:
\begin{enumerate}
    \item How can the structural duality of the orbits $n$ in Bohr's model (in Fig. \ref{fig:structure}) be extended to low-inertia power systems?
    \item With increased penetration from CIGs, $m_i$ from SMs will be sequentially depleted that will then displace the space coordinates of the center of mass $\mathbf{R}$. How will this displacement affect a shift in the coordinates of the orbits around it? 
\end{enumerate}
{To prove this, we will now use the Huygens-Steiner theorem \cite{buchho}, which states that the moment of inertia of a mechanical system with respect to the sum of the moment of inertia with respect to the coordinates for the center of mass $\mathbf{R}$ and the product of the component mass $m_e$ at the desired location by the square distance between the two points $r_1$ (see the distance OA between nucleus and orbit 1 in Fig. 2):
\begin{eqnarray}
J = J_M + m_e r_1^2
\end{eqnarray}
Hence, the difference in inertia $\Delta J$ between OA can then be simplified using (1) by multiplying $\omega \od[2]{\delta}{t}$ in both sides to obtain:
\begin{equation}
    P_m - P_e = m_e r_1^2 \omega \od[2]{\delta}{t}
\end{equation}
Equivalencing the LHS as the total transmitted power between the two segments OA, we can obtain:
\begin{eqnarray}
    V_o V_a B_{oa} \text{sin} \ \delta = C r^2_1 \omega \od[2]{\delta}{t}
\end{eqnarray}
where, $V_o$ and $V_a$ represent the voltages at point O and A, respectively. Furthermore, $B_{oa}$ represent the total susceptance between OA. Considering close to unit voltages in (6) with fixed values of $\omega$ and the characteristic mass $m_e$ of VSG at point A (given by the value $C$ in Remark II), (6) can be reduced to small-angle approximations using:
\begin{equation}
    \kappa (\text{sin} \ \delta - \od[2]{\delta}{t}) = 0
\end{equation}
with $\kappa$ = $\frac{B_{oa}}{C \omega}$, only if, $r_1$ = $\sqrt{\frac{B_{oa}}{C\omega}}$. Hence, we establish from this proof that the orbital radius is directly dependent on the equivalent impedance from the center of mass to different VSGs.

\textit{\textbf{Corollary II:} The radius $r_i$ in the atomic model corresponds to the equivalent electrical impedance between the buses representing the nucleus and orbits. In short, they are analogous to the quantization levels for different orbits}.

It not only aligns with Bohr's third postulate, but also answers the energy dissipation theory in the atomic model along electrical path for power systems, corresponding to Remark III.

We will now test our hypothesis and answer the abovementioned queries using different case studies of penetration from multiple VSGs and SMs to provide new insights the anatomical modeling co-relationship in the next section.
\begin{figure}[!t]
\centering
\includegraphics[width=\columnwidth]{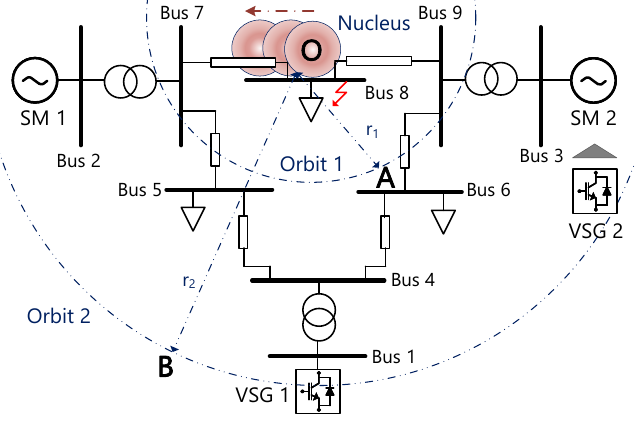}
\caption{Anatomical model of IEEE 9-bus system \cite{psc} -- with increase in number of VSGs, the center of mass (significant reduction in system mass) displaces close to SMs.}
\label{fig:atommodel}
\end{figure}

\begin{figure}[!b]
\centering
\includegraphics[width=0.8\columnwidth]{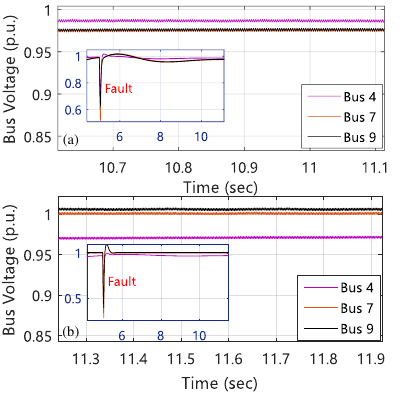}
\caption{Simulation results of the voltages at bus 4, 7 \& 9 in IEEE 9-bus system after clearance of fault in bus 8 for: (a) case II, (b) case IV -- corollary III is established with a clear illustration of the orbits for VSGs and nucleus always attained at a discrete distance of large quantum numbers.}
\label{fig:result}
\end{figure}
\begin{figure}[!h]
\centering
\includegraphics[width=0.85\columnwidth]{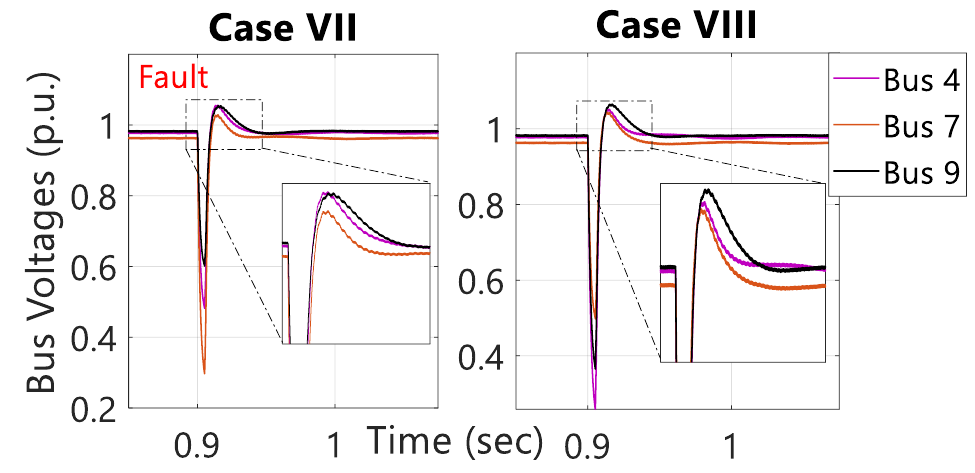}
\caption{Simulation results of the voltages at bus 4, 7 \& 9 in IEEE 9-bus system after clearance of fault in bus 8 for: (a) Case VII with homogeneous parameters for VSGs, (b) case VIII with heterogeneous parameters for VSGs -- it establishes that the heterogeneous $J$ further deviates from the center of mass comprising a significant distribution of kinetic energy leading to more oscillations and less damping.}
\label{fig:newresult}
\end{figure}

\section{Structural Insights}
{To test our hypothesis}, we consider different combinations of VSGs and SMs in IEEE 9-bus system using PSCAD simulation environment. For IEEE 9-bus system with 3 generators, we consider three case studies in Table \ref{tab:equivalence} with: 1) 1 VSG and 2 SMs, 2) 2 VSGs and 1 SM, and 3) 3 VSGs/SMs with homogeneous/heterogeneous parameters. As shown in Fig. \ref{fig:atommodel}, the center of mass is located at the center due to the placement of SM 1 and SM 2 at both ends (Refer to Case I in Table \ref{tab:equivalence}). In Case I after the clearance of the fault in bus 8, it can be observed that the corresponding voltages across the buses (Bus 3 \& 1) with SM are attained at a discrete difference from the bus (Bus 2) with VSG. Furthermore, the quantization of the difference can also be seen in the power generation from each bus. It should be worthy notifying that the voltage difference between both the SM buses in Case I-III (in Table I) is expected to be within a tolerable range of 0.005 p.u in a distributed particle system. In Case II, when VSG1 is integrated into bus 6, a similar pattern is observed where its bus voltage and power is attained at a discrete difference of large quantum numbers from bus 1 \& 3. A time-domain simulation result corresponding to Case II is also shown in Fig. \ref{fig:result}(a) after the fault is cleared. In Case III, the placement of generators are shuffled with SMs integrated into bus 2 and 3 and VSG in bus 1. Although this changes the coordinates of the center of mass $\mathbf{R}$, similar behavior is observed with the difference in bus voltages and generated power.

\textit{\textbf{Corollary III:} From the discussion above with SMs and VSG representing the nucleus and electrons respectively, terminal voltages can be adjudged as the structural duality of orbits $n$ in low-inertia power systems for absorption/dissipation of a given activation energy $\Delta E$.}


As shown in Fig. \ref{fig:atommodel}, when the number of VSGs are increased (with VSG 2 replaced with SM 2 in bus 3), the center of mass $\mathbf{R}$ is displaced towards the bus closest to SM 1 as per corollary I with a significant reduction in system mass. However, following from the cases I-V in Table \ref{tab:equivalence}, the difference of voltages across SM and VSG is not consistent due to different activation energies. Using Remark III and corollary III, we can conclude that a viable definition of $r_i$ in low-inertia power systems can be the equivalent electrical impedance between the bus representing the nucleus (SM, as per corollary I) and the orbits with electrons (VSGs). In IEEE 9-bus system, this can be validated using case IV (see Fig. 3(b)) \& V, wherein the placement of SM and VSGs are shuffled by integrating them into different buses. Although they are aligned with corollary III, the electrical radius $r_i$ affects their corresponding voltage profiles.
\begin{table}[!thb]
\centering
\caption{VSG Parameters for Case VII and VIII}
\begin{tabular}{@{\extracolsep{\fill}}cccc}
\toprule
  & \textbf{VSG 1} & \textbf{VSG 2} & \textbf{VSG 3}\\
  \midrule
      Rating (MW) & 150 & 150 & 150\\
      Inertia co-efficient ($J$) & 2.2 & \begin{tabular}{@{}c@{}}2.2 \\ 1.1\end{tabular} & \begin{tabular}{@{}c@{}}2.2* \\ 1.1\end{tabular}\\
      Damping co-efficient ($D$) & 2.5 & \begin{tabular}{@{}c@{}}2.5 \\ 1.5\end{tabular} & \begin{tabular}{@{}c@{}}2.5 \\ 1.5\end{tabular}\\
      Voltage loop gain & 5 & 5 & 5\\
      Time constant & 0.2 & 0.2 & 0.2\\
\bottomrule
\end{tabular}\\
\footnotesize{$^*$ denote the two heterogeneous scenarios of VSGs.}
\label{tab:vsg}
\end{table}

{We then analyze the baseline comparison in a system only comprising of all SMs \& VSGs with homogeneous and heterogeneous parameters in Case VI-VIII in Table I. The VSG parameters used for the simulation are provided in Table II. Based on the results, we can extend the translation of electrical distance $r_i$ affecting the corresponding voltage profiles accordingly, thereby confirming corollary III.}

{On the other hand, we also conducted further analysis in Fig. 4 on how heterogeneity of VSG parameters can be translated to our postulates. To investigate the system response in Case VII and VIII, the voltage oscillations have increased post-fault clearance owing to the reduction in overall system mass as compared to the significantly larger mass in Case VI. However, when heterogeneous inertia and damping coefficients are considered in case VIII, the oscillations have increased in Fig. 4 with further reduction in the system inertia from Case VII, leading to reduced damping.}

\section{Conclusions and Future Scope of Work}
This letter determines a fundamental atomic model with clearly defined structural corollaries between Bohr's model and low-inertia power systems. With the preliminary anatomical model intact, we aim to explore a detailed modeling framework and stability using atomic physics based principles as a future extension of this work. Furthermore, detailed theoretical proof on the duality criteria of frequency in Fig. 1 with its rationale will be provided.


\bibliographystyle{IEEEtran}
\bibliography{refs}

\newpage

 




\vfill

\end{document}